\documentclass[sn-mathphys-num]{sn-jnl}

\usepackage{graphicx}%
\usepackage{epstopdf}
\usepackage{multirow}%
\usepackage{amsmath,amssymb,amsfonts}%
\usepackage{amsthm}%
\usepackage{mathrsfs}%
\usepackage[title]{appendix}%
\usepackage{xcolor}%
\usepackage{textcomp}%
\usepackage{manyfoot}%
\usepackage{booktabs}%
\usepackage{algorithm}%
\usepackage{algorithmicx}%
\usepackage{algpseudocode}%
\usepackage{listings}%

\theoremstyle{thmstyleone}%
\theoremstyle{thmstyletwo}%

\theoremstyle{thmstylethree}%

\begin{document}

\title[Title]{Formulation of a  {one-dimensional} electrostatic plasma model for testing the validity of kinetic theory}


\author*[1]{\fnm{F.} \sur{Pegoraro}}\email{francesco.pegoraro@unipi.it}

\author[2]{\fnm{P. J.} \sur{Morrison}}\email{morrison@physics.utexas.edu}
\equalcont{These authors contributed equally to this work.}

\author[1,3]{\fnm{D.} \sur{Manzini}}\email{davide.manzini@lpp.polytechnique.fr}
\equalcont{These authors contributed equally to this work.}

\author[1]{\fnm{F.} \sur{Califano}}\email{francesco.califano@unipi.it}
\equalcont{These authors contributed equally to this work.}

\affil*[1]{\orgdiv{Physics Department}, \orgname{University of Pisa}, 
\city{Pisa}, 
\country{Italy}}

\affil[2]{\orgdiv{Physics Department and Institute for Fusion Studies}, \orgname{University of Texas at Austin},
\city{Austin},
\state{TX}, \country{USA}}

\affil[3]{\orgdiv{Laboratoire de Physique des Plasmas}, \orgname{ \'Ecole Polytechnique, Sorbonne Universit\'e}, 
 \city{Paris}, 
 \country{France}}


\abstract{
We present a one-dimensional (1-D) model composed of aligned,  electrostatically interacting charged disks,  conceived in order to address in a computable model the validity of the Bogoliubov assumption on the decay of  particle correlations in the  
Born-Bogoliubov-Green-Kirkwood-Yvon hierarchy.  This assumption is  a basic premise of plasma kinetic theory.  
The disk model exhibits spatially 1-D features at short distances,  but retains  3-D  features  at large distances.  Here  the collective dynamics  of this model plasma is investigated by solving  the corresponding Vlasov equation.  In  addition, the  implementation  of the model for the numerical validation of the Bogoliubov assumption is  formulated. }




\maketitle


%
%
%
%
 

%

\section{Introduction}\label{intr}

The  Born-Bogoliubov-Green-Kirkwood-Yvon (BBGKY) hierarchy of  phase-space  equations is the  starting block \cite{Cercignani}  in the description  of a wide class of multi-particle interacting systems,  ranging from stellar systems \cite{Chav}, to electromagnetic  plasmas, in  both  the classical  \cite{Ichi, Kampen} and relativistic    \cite{Vereshn}  regimes,  to Quantum ElectroDynamics  plasmas  \cite{Fed},  down to ultracold systems \cite{Kronk, cross},   and to quantum spin dynamics \cite{Pucci}.   A major difficulty   is related to the need to construct  a  consistent closure of the infinite  chain  of equations of  the  BBGKY  hierarchy.

The problem  of dealing with this infinite chain  has been addressed in the literature with  different approaches  and under different  conditions,  both from the mathematical and from the physical viewpoints.  
For example, in  \cite{Ryabu}  a regularized   representation  of the  solution  of the BBGKY hierarchy for a 1-D infinite system of hard spheres was proven for given initial data, while  \cite{Gubal}  addressed  the Cauchy problem for   an infinite 1-D  particle system, where the particles interact with each other by a finite-range pair potential with a hard core.   In   \cite{Baal} a  closure of the BBGKY hierarchy was  developed by enforcing  an exact equilibrium limit at all orders.  This leads  to a convergent kinetic equation that can extend plasma kinetic theory into the regime of strong Coulomb coupling. In  \cite{Singleton}   dimensional reduction  and  dimensional continuation techniques allowed  the adoption of  a regularization scheme of the spatial divergences  that occur both at short and at long distances in a Coulomb plasma.

While most of the investigations on the solution or the  truncation of the BBGKY hierarchy  are restricted to spatially homogeneous systems, in   \cite{Chav}   the growth of correlations  with time  was  studied in the context of  the relaxation of self-gravitating systems and  a  general kinetic equation was  derived  that can be applied to spatially inhomogeneous systems, including ones 
 that take into account memory effects that cannot be accounted for if exact equilibrium conditions are  imposed.  
With a different perspective  in  \cite{Morrison} it was shown that the  equations for the evolution of the
$i$-point functions in the BBGKY hierarchy  can be cast in  Hamiltonian form  with  Lie-Poisson
brackets that involve  a Lie algebra  constructed from the algebra of $n$-point functions.  It was  stated  that this structure can be  inherited by truncated subsystems in the hierarchy.

As shown above, most of these investigations rely on the assumption of  a time scale separation, the so-called Bogoliubov assumption \cite{Bogo} (see also   \cite{Kampen}), where the particle correlations decay on a fast  time scale  and  reach an asymptotic form that is uniquely determined by the instantaneous form of the uncorrelated functions.

The aim of the present  article is to provide a means for testing  this assumption numerically, in the case of the 2-point function of a Coulomb plasma, by retaining in the hierarchy only contributions of order ${\cal{O} }(g)$, where $g\ll 1$ is the standard plasma parameter.  

It must however be remarked that,  at present, such a numerical  test is  prohibitive in the case of  a 3-D system,  since  the equation for the time evolution of the two-point function would be 13-dimensional (one time dimension plus the 12 dimensions of the two-particle phase space).  In 1-D  the equation for the two-point distribution would be 5-dimensional (time plus  the 4-dimensions of the two-particle phase space). However in 1-D the electric field generated by a one-dimensional charge (a charge foil)  does not
decay with distance so that the two-foil interaction energy diverges at infinity. This makes it unclear how to define a finite correlation length and  to define binary collisions between charge foils.   A reasonable compromise is to introduce an interaction potential that behaves as a 1-D system at short distances while it decays as the inverse of the distance at large distances. It must be noted that  any interaction potential of this type introduces a length scale, independent of the plasma parameters such as temperature and density, that is not present in the Coulomb interaction. This length will appear explicitly in the collective plasma dynamics.

 As a specific example of such a potential, here  we  consider  \cite{wtw} a 1-D plasma made of aligned, uniformly
charged disks (see Figure f{CP0}),  where the characteristic  scale length is the disk radius $b$.  In this case the electrostatic interaction energy between two charged disks is finite both at zero and at infinite distance 
from the source and its absolute value is a monotonically decreasing function of 
the distance.
 
\begin{figure} [h!]\center
\includegraphics[width=0.7\textwidth]{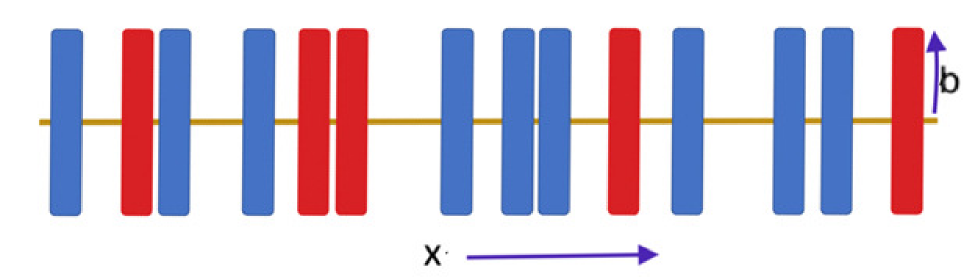}
\caption{\footnotesize  Plasma of two species of oppositely charged   disks: blue positively charged, red negatively charged disks.  The disks are aligned and can only move in the $x$ direction {,} i.e. along their axis,  but can pass each other.}
\label{CP0}
\end{figure}

This  article is organized as follows.  In Section  f{BBGKY}  the basic equations of the BBGKY hierarchy are briefly recalled, together with  the  formulation of the Bogoliubov assumption.  In Section  f{model} the electrostatic properties  of the disk model are described  and some useful expressions are listed.  In  Section  f{dynam} the collective disk plasma dynamics is investigated  by solving the corresponding Vlasov equation  in order to clarify where and how the dynamics of such a model system can
mimic the dynamics of a Coulomb plasma, at least as long as the longitudinal electric field limit is concerned. Finally in  Section  f{Numer}  the  system of  equations to be solved  numerically is derived in explicit form  and  in Section   f{fut} the numerical approach that will be applied is briefly described.

\section{The BBGKY  {hierarchy} and Bogoliubov's assumption}
\label{BBGKY}

The time evolution of  a system of $N$ classical particles obeys the Liouville equation for the  $N$-point distribution  function in $6N$-dimensional phase space (3 space and 3 momentum coordinates per particle) 
\begin{equation}\label{BB0}  
F^{(N)} =  F^{(N)}({\bf q}_1...{\bf q}_N , {\bf p}_1...{\bf p_N},t)\,, 
\end{equation}
that  can be written as 
\begin{equation}\label{BB1}
\frac{\partial   F^{(N)}}{\partial t}  +  \sum_{i=1}^N  \frac{{\bf p}_i}{m_i} \cdot \frac{\partial F^{(N)}}{\partial{\bf q}_i}
+\sum_{i=1}^N {\cal F}_i \cdot \frac{\partial F^{(N)}}{\partial{\bf p}_i} = 0\, , 
\end{equation}
with ${\bf q}_i,{\bf p}_i$ the coordinate and momentum of the $i^{th}$ particle with mass $m_i$, and  ${\cal F}_i$  the total force acting on the $i^{th}$   particle, 
\begin{equation}\label{BB2}
 {\cal F}_{i } =  \sum_{j=1, j\not= i}^N  {\cal F}_{i,j } =-  \sum_{j=1, j\not= i}^N \frac{\partial V_{ij}}{\partial {\bf q}_i}\,,    
 \end{equation} 
 with $V_{ij} $ the  interaction  potential,
and
  \begin{equation}\label{BB3b}
  \int d{\bf q}_{1}.. d{\bf q}_N  {\,}   d{\bf p}_{1}.. d{\bf p}_N \,
 \,   \, F^{(N)}({\bf q}_1...{\bf q}_N  {\,}  {\bf p}_1...{\bf p_N},t)   =1\, ,  
 \end{equation} 
where the spatial integrations are over a finite volume.

By integration over the variables of a subset of particles, the Liouville equation can be transformed into a chain of equations for the $s$-point distribution function
\begin{equation}\label{BB3}
 F^{(s)}({\bf q}_1...{\bf q}_s  {\,} {\bf p}_1...{\bf p_s},t)= \int d{\bf q}_{s+1}.. d{\bf q}_N  {\,}  d{\bf p}_{s+1}.. d{\bf p}_N \,
 \,    F^{(N)}({\bf q}_1...{\bf q}_N , {\bf p}_1...{\bf p_N},t) \,,
   \end{equation}
that read as 
 \begin{align}\label{BB4} 
\frac{ \partial F^{(s)}}{\partial t} +  \sum_{i=1}^s  \frac{{\bf p}_i}{m_i} \cdot  \frac{\partial F^{(s)}}{\partial{\bf q}_i}
&+\sum_{i=1}^s  \sum_{j=1, j\not= i}^s {\cal F}_{i, j} \cdot  \frac{\partial F^{(s)}}{\partial{\bf p}_i} 
\nonumber\\
&=    - (N-s)  \sum_{i=1}^s \int  {\cal F}_{i ,s+1} \cdot \frac{\partial F^{(s +1)}}{\partial{\bf p}_i}  d{\bf q}_{s+1} d{\bf p}_{s+1}\,.
 \end{align}
This system of equations is not closed:  the first equation connects the evolution of the one-particle distribution function $ F^{(1)}$  with the two-particle distribution function $ F^{(2)}$,  the second equation connects the two-particle distribution function  $ F^{(2)}$ with the three-particle distribution  function $ F^{(3)}$ and so on.

For  the case of the one particle  distribution function ($s = 1$)  we obtain  
 \begin{equation}\label{BB5} 
\frac{ \partial   {F^{(1)}}}{\partial t} +  \ \frac{{\bf p}_1}{m_1} \cdot  \frac{\partial   {F^{(1)}}}{\partial{\bf q}_1}
=  - (N-1)  \ \int  {\cal F}_{1 ,2} \cdot \frac{\partial   {F^{(2)}}}{\partial{\bf p}_1}  d{\bf q}_{2} d{\bf p}_{2}\,,
 \end{equation}  
where  $F^{(2)}({\bf q}_{1}, {\bf  q}_{2},  {\bf p}_{1} , {\bf p}_{2},t)$  is linked to  $F^{(3)}({\bf q}_{1}, {\bf  q}_{2},  {\bf  q}_{3}, {\bf p}_{1} , {\bf p}_{2}, {\bf  p}_{3},t)$ 
 by
 \begin{align}\label{BB6}  
\frac{ \partial   {F^{(2)}}}{\partial t} +   \frac{{\bf p}_1}{m_1} \cdot  \frac{\partial   {F^{(2)}}}{\partial{\bf q}_1}
&+\ \frac{{\bf p}_2}{m_2} \cdot  \frac{\partial   {F^{(2)})}}{\partial{\bf q}_2} + {\cal F}_{1,2}\cdot  \frac{\partial  {F^{(2)}}}{\partial{\bf p}_1} +  {\cal F}_{2,1}\cdot  \frac{\partial   {F^{(2)}}}{\partial{\bf p}_2}   
\nonumber \\ 
&=
- (N-2)   \int \left[ 
{\cal F}_{1,3} \cdot \frac{\partial  {F^{(3)}}} {\partial{\bf p}_1}  + {\cal F}_{2,3}\cdot  \frac{\partial {F^{(3)}}}{\partial{\bf p}_2}
\right] d{\bf q}_{3}  d{\bf p}_{3}\,. 
  \end{align}
 Integration of ( f{BB6}) over ${\bf  q}_{2}$ and ${\bf  p}_{2}$ returns  ( f{BB5}). The space integration is over a finite volume and all $F^{(s)}$  are normalized to one, i.e., 
 \[
 \int d{\bf q}_{1}.. d{\bf q}_s,   d{\bf p}_{1}.. d{\bf p}_s \,
 F^{(s)}({\bf q}_1...{\bf q}_s , {\bf p}_1...{\bf p_s},t) = 1\,.
 \]

As is standard \cite{Kampen,Krall},  we adopt an expansion in terms of the plasma parameter $g\ll 1$ (defined as the inverse of the number of particles in a Debye sphere) and write 
  \begin{equation}\label{BB7}  
  {F^{(2)}} ( {\bf q}_{1}, {\bf  q}_{2},   {\bf p}_{1} , {\bf p}_{2},t) =   {F^{(1)}} ({\bf q}_{1}, {\bf p}_{1},t)   {F^{(1)}}(  {\bf  q}_{2}, {\bf p}_{2},t)   +   {\Delta  F^{(2)} }({\bf q}_{1},{\bf p}_{1}, {\bf  q}_{2}, {\bf p}_{2},t)  \,,
 \end{equation}  
 where $ {\Delta  F^{(2)} }({\bf q}_{1},{\bf p}_{1}, {\bf  q}_{2}, {\bf p}_{2},t)$ is the correlated part, and similarly 
  \begin{align}\label{BB8}  
&{F^{(3)}} ( {\bf q}_{1}, {\bf  q}_{2},  {\bf  q}_{3}, {\bf p}_{1} , {\bf p}_{2}, {\bf  p}_{3},t) =   {F^{(1)}} ({\bf q}_{1}, {\bf p}_{1},t)   {F^{(1)}}(  {\bf  q}_{2}, {\bf p}_{2},t)  {F^{(1)}}({\bf  q}_{3}, {\bf p}_{3},t)\,\,    
\nonumber \\
& \hspace{.5cm} +\  {F^{(1)}}({\bf q}_{1}, {\bf p}_{1},t) \,    {\Delta  F^{(2)} } ({\bf q}_{2},{\bf p}_{2}, {\bf  q}_{3}, {\bf p}_{3},t) \   
 + {F^{(1)}} ({\bf q}_{2}, {\bf p}_{2},t)  \,   {\Delta  F^{(2)} } ({\bf q}_{3},{\bf p}_{3}, {\bf  q}_{1}, {\bf p}_{1},t)     
 \nonumber  \\
 &\hspace{1.5cm} +   {F^{(1)}} ({\bf q}_{3}, {\bf p}_{3},t) \,    {\Delta  F^{(2)} } ({\bf q}_{1},{\bf p}_{1}, {\bf  q}_{2}, {\bf p}_{2},t)  \   +  \     {\cal O} (g^2) \,.  
   \end{align}  
This  expansion leads to a closed system of evolution equations for  the correlated part of the 2-point distribution function $\Delta  F^{(2)}$.

The basic step  in the  Bogolyubov's method  of solution of the above system of equations is  the assumption that    $ \Delta F^{(2)}
({\bf q}_{1}, {\bf p}_{1},   {\bf  q}_{2}, {\bf  p}_{2},t)   $
rapidly reaches an asymptotic form ${\tilde  \Delta F^{(2)}} ({\bf q}_{1}, {\bf p}_{1},   {\bf  q}_{2}, {\bf  p}_{2},t) $ that   is uniquely determined by the instantaneous form  of the function $F^{(1)}( {\bf  q}, {\bf p}, t)$    (time scale separation), i.e.,  at each  time  $t$  we have 
   \begin{equation}\label{BB9}   
   {\tilde  \Delta F^{(2)}}({\bf q}_{1}, {\bf p}_{1},   {\bf  q}_{2}, {\bf  p}_{2},t) = {\tilde  \Delta F^{(2)}}({\bf q}_{1}, {\bf p}_{1},   {\bf  q}_{2}, {\bf p}_{2}| F^{(1)}( {\bf  q}, {\bf p}, t))\,.
   \end{equation}
  When  ( f{BB9}) is substituted into  \eqref{BB5} and  \eqref{BB7}, a closed equation for $F^{(1)}$ is obtained  \cite{Kampen} that 
 is the basis for the derivation of the collision operators of  the Vlasov-Landau-Lenard-Balescu  \cite{landau,lenard,balescu} theories.

\section{The disk-disk interaction} \label{model}

   { As  indicated by \eqref{BB2} a central feature   in the BBGKY hierarchy is the interparticle potential  $V_{ij}$ as it determines their  interaction range and finally the dimensionless  expansion  parameter that can be used  in order to truncate the  BBGKY hierarchy  and  construct a closed set of equations on which the Bogoliubov assumption can be   tested.  
\\ As already mentioned in the Introduction, in order to reduce  the  dimensionality  of the system of equations to be solved numerically,  we  resort  to a model one-dimensional interaction potential regularized at large distances, as is the case of the interaction potential between aligned charged disks.  Therefore in}  this and in the next  section  the main properties of  the disk-disk interaction and of the collective disk plasma  dynamics are described with the aim of identifying  which properties  correspond to and which differ  from those of  a  {three-dimensional} Coulomb plasma \cite{arxiv}.  {This point will be readdressed  at the end of Sec. f{Numer} in a comment on   the relevance of a one-dimensional model to the validation of the Bogoliubov assumption in a three-dimensional plasma.}
\\
   The  interaction potential  between aligned  disks is known in explicit form \cite{Disk} but, more generally,  the interaction potential could be assigned without referring to  uniformly charged disks  in a form that is regular at short
distances and decays monotonically  at large distances  as the inverse of the distance.

The interaction energy ${\cal W} (x_1,x_2)$  between two uniformly  charged  infinitely thin disks  of radius $b$ located  at $x_1$ and $x_2$,  with charge $Q_1,$ and $Q_2$, respectively,  can be written as  (see Appendix  f{App1}),
 \begin{equation}\label{1}  
{\cal W } (x_1,x_2) =  4  \frac{Q_1\,  Q_2}{b} V(\xi) = Q_1 \varphi_2 =  Q_2 \varphi_1\,,
\end{equation}
where
\begin{equation} \label{2} 
V(\xi) = - \frac{\xi}{2} \left\{  1 - \frac{1}{3\pi} \left[ (4 -\xi^2) E(-4/\xi^2)  +  (4 +\xi^2) K(-4/\xi^2) \right]   \right   \},
\end{equation}
and $\varphi_1$, $\varphi_2$ are the potentials generated by the disks $1$ and $2$, respectively. Here $\xi(x_1,x_2) = |x_1 - x_2|/b$,  $ Q_1=\pi b^2\sigma_1$ and $ Q_2 = \pi b^2\sigma_2$ are the (fixed) charges  of the two disks, $\sigma_{1,2}$ are their surface density and   $ E(-4/\xi^2) $  and $ K(-4/\xi^2) $  are elliptic integrals  \cite{abram, Nist}.  
A plot of $V(\xi)$ is given in Figure  f{figV}.

\begin{figure} [h!]\center
\includegraphics[width=0.6\textwidth]{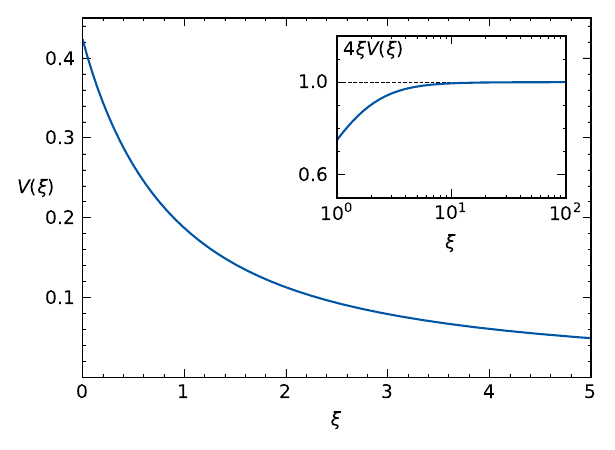}
\caption{Plot of the potential $V(\xi) $ versus  the normalized distance $\xi$.    Note the $1/(4\xi)$-behavior for large $\xi$   {shown in the insert} and the finite value $4/(3 \pi)$ at $\xi = 0$.}
\label{figV}
\end{figure}

The electric force ${\cal F}_{Q_2}$ acting on disk 1  due to disk 2 (see Figure  f{figV'})  is given by
\begin{equation}  \label{3} 
 {\cal F}_{Q_2} (x_1) = -  {\cal F}_{Q_1} (x_2)=  -(4  Q_1 Q_2/b^2) \,  V'(\xi) \, {\rm sign} \,(x_1 -x_2) , 
\end{equation} 
with 
\begin{equation}  \label{3v} V'(\xi) =\frac{d V(\xi)}{ d\xi }=  \frac{1}{2}  +  \frac{1}{2\pi} \left[ \xi^2 E(-4/\xi^2)  -  (4 +\xi^2) K(-4/\xi^2) \right]   \,.
\end{equation}
The Fourier transform of $V(\xi)$ (see Appendix  f{App2}) is given by 
\begin{align} \label{Vfur}
  {\hat V} (kb) &=   \int _{-\infty}^{+\infty} \frac{dx}{b} \,V(|x|/b) \exp{-[i (kb) \, (x/b)]}     
 \nonumber \\ 
  &=
2  \int_{0}^{+\infty} \!\!ds \, [J_1(s)/s]^2  \,  \frac{s}{(kb)^2 + s^2} = \frac{1}{  \pi^{1/2} (kb)^2}G^{2,2}_{2,4}\left((kb)^2\left| ^{1/2,1}_{1,1,-1,0} \right.\right) ,
\end{align} 
 where $G^{2,2}_{2,4}\left((kb)^2\left| ^{1/2,1}_{1,1,-1,0} \right. \right)$ is the Meijer G function \cite{abram, Nist}.

\begin{figure} [h!] \center 
\includegraphics[width=0.6\textwidth]{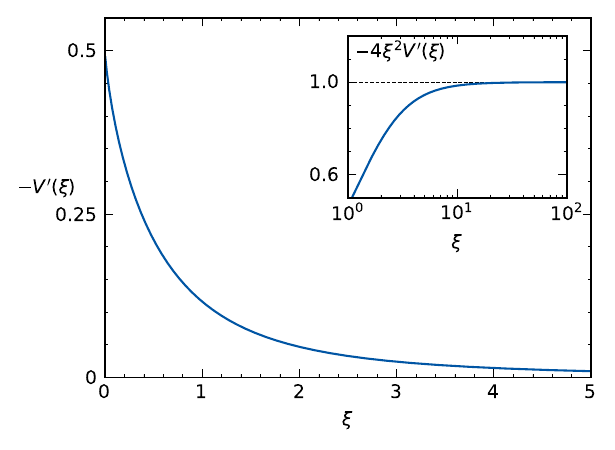}
\caption{Plot of  $-V'(\xi) $ versus $\xi$.\,   Note the $1/(4\xi^2)$ behavior for large $\xi$   {shown in the insert, and the value $-V'(0) = 1/2$}.}
\label{figV'}
\end{figure}

In the limits of long wavelengths  with respect to the disk radius $b$, i.e., $(kb)^2\ll  1 $, we find  from  ( f{Vfur}) 
 \begin{align}  \label{Vexp}  
{  {\hat V} (kb) } &=     (1/8)\left\{ 5 - 6 \gamma - 2 \ln{[(kb)^2]} - 2 \psi(3/2) \right\}
 \nonumber \\ & + (1/48)\left\{ 13 - 9 \gamma - 3 \ln{[(kb)^2]} - 3 \psi(5/2) \right\} (kb)^2   + \dots\,,  
\end{align} 
with $\gamma = 0.577216$ {,  $\psi (z) = d \ln{\Gamma(z)}/dz$ and $\Gamma(z)$  the Gamma function \cite{abram, Nist}}.  For short wavelengths, i.e.,   $(kb)^2\gg  1$,  we find
\begin{equation}        
{ {\hat V} (kb) }=  { \frac{1}{   (kb)^2} }   + \dots\,.
\end{equation} 
For  short wavelengths  we recover the Coulomb spectrum  ${\hat V} \propto 1/k^2$ while  for small $kb$ the disk and Coulomb  spectra differ.   A plot of ${ {\hat V} (kb) }$  is given in Figure  f{fur}.  The logarithmic  behavior  at small wave-numbers  does not depend on the specific  model adopted and  arises  from the  imposed $1/\xi $ dependence of the interaction potential  at large distances in a 1-D configuration.

\begin{figure} [h!] 
\includegraphics[width=0.5\textwidth]{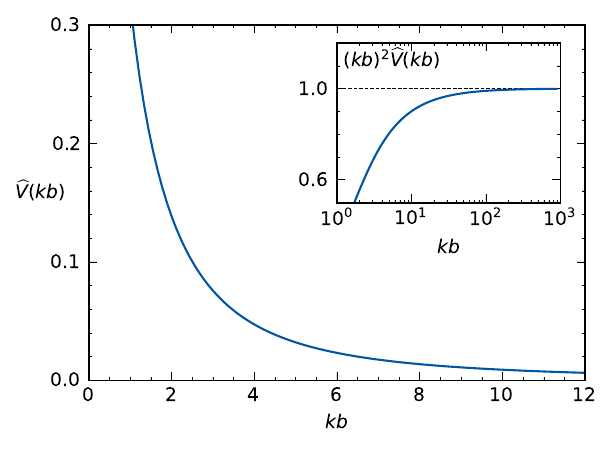}
\includegraphics[width=0.49\textwidth]{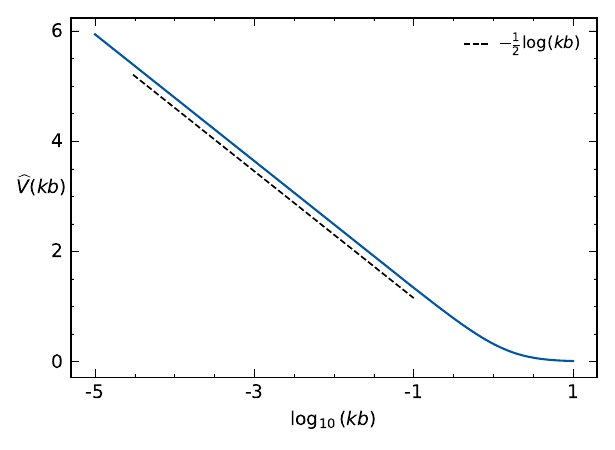},
\caption{Plot of ${\hat V}(kb)$  versus  $kb$.  { The insert  in the left frame shows  the  $1/(kb)^2 $ for  large $kb$, 
 while the right frame displays the logarithmic behaviour as $kb \to 0$
 }}
\label{G-small}
\end{figure} 

\section{Collective disk plasma dynamics} \label{dynam}

 {In this section the interaction potential described in Sec. f{model} is used in order to define  the characteristic  dynamical time scales and lengths of a one-dimensional  disk plasma  (Sec. f{dynamcold}, f{CDPW})  and to compare them with the corresponding quantities  in a Coulomb three-dimensional plasma. In Secs. f{screening},   f{g-Parameter} the expressions of the screening length  and of the dimensionless expansion   parameter  that will be used for truncating the   BBGKY hierarchy   in a disk plasma are  derived. In Sec  f{dynamwarm} we  consider the disk Vlasov equation which in the present context corresponds to assuming that the disks in the plasma are  and remain uncorrelated.  Solving the disk Vlasov equation serves a  dual purpose. First it allows to confront  the wave spectra  in  a disk  plasma  and in a three-dimensional Coulomb plasma and, through the wave group velocity, the speed of the  transmission of information through the plasma.   Then, it will allow us to  compare the results obtained  for the disk plasma excitations  in the uncorrelated Vlasov case with those  that will be found in the case when  the correlations  are retained and explicitly solved for, i.e. with  the solutions  that  will be obtained by integrating numerically equations    \eqref{f1} and    \eqref{f1b}    of Sec. f{closed}.}

We consider a globally neutral disk plasma and, in analogy with a Coulomb plasma, take the continuous  density limit.
We introduce positive and negative charged disk densities and  a  collective (mean) disk electric field.  The mean electric field is  calculated from the total  disk charge density $\rho$  and  a Green's function with a kernel that involves the interaction electrostatic potential  $V(\xi)$  of  \eqref{2},  which plays the role of the Coulomb potential in a 3-D plasma.   
\subsection{Cold fluid equations} \label{dynamcold}

Let  $n_e(x,t)$ be the electron disk number density   (number of  negatively charged disks per unit length) and $n_i(x,t)$ be the ion disk number density  (number of  positively charged disks per unit length).   For simplicity,  in this section,  we  assume the charges on the disks  satisfy  $Q = Q_i= -Q_e$, so that  the  total disk-charge density  is given by 
\begin{equation} \label{4}
\rho(x) = Q \, [n_i -  n_e(x,t)]\,,
\end{equation}
where $n_i$ is constant.  The electron disk  density satisfies the continuity equation
\begin{equation} \label{5}
\partial_t n_e(x,t) + \partial_x [n_e(x,t) \, u_e(x,t) ] = 0,
\end{equation}
where $u_e(x,t)$  is the electron disk fluid velocity.

The  electrostatic potential  created by the  disk charge density  can be written in convolution integral form as 
\begin{equation} \label{6}
\varphi(x)  =  \frac{4}{b} \int_{-\infty}^{+\infty} \!\!dx' \, \rho(x')V(|x-x'|/b)\,,  
\end{equation}
which  for $\rho(x')  = Q\, \delta (x' -x_1)$ returns  the potential generated by a single disk.
 The corresponding  electric force acting on a disk of charge $Q_j$ \, ($j= i,e$)   can be written as 
\begin{equation} \label{7}
{\cal F}_j(x)  =  -\frac{4 Q_j}{b} \frac {d}{dx} \int_{-\infty}^{+\infty}\!\! dx' \, \rho(x')V(|x-x'|/b)\,.  
\end{equation}
To close the system we  may supplement the continuity equation  with the   momentum equation of a ``cold''  disk plasma
\begin{equation} \label{14a}
M_e n_e\left( \partial_t + u_e \partial_x\right) u_e =  n_e {\cal F}_e\,,  
\end{equation}
where here and in the following the  electron and ion disk masses are denoted by  $M_{e.i}$.

\subsection{Cold plasma disk waves}\label{CDPW}

We consider the case where the disk thermal motion can be neglected  with respect to the wave phase velocity and derive the linear dispersion relation of the cold disk plasma.
Linearization  of  \eqref{5} and  \eqref{14a}  for a monochromatic  wave, 
\begin{equation} \label{5b} 
 {\tilde n}_{e}(x,t) = {\tilde n}_{0}  \exp {[-i (\omega t -  kx)]} \quad \mathrm{and}\quad
{\tilde u}_e(x,t) =  {\tilde u}_0\, \exp {[-i (\omega t -  kx)]}\,, 
 \end{equation}
in a stationary  homogeneous equilibrium with density $n_0$  and immobile ion disks yields 
\begin{equation} \label{15}
{\tilde u_e} = \frac{{\tilde n}_e}{n_0}  \frac{\omega}{ k} \quad\mathrm{and}\quad 
 {\tilde u_e} 
= \frac{ 4 k {\tilde n}_e}{M_e  \omega} \, Q^2 {\hat {V}}(kb) \,,  
\end{equation}
which give
\begin{equation} \label{16}
\omega^2 = \omega_{Dpe}^2\, (k b)^2 \,{\hat {V}}(kb)\,,  
\quad {\rm where} \quad \omega_{Dpe}^2= \frac {4 Q^2 n_0}{M_e b^2}\,. 
\end{equation}
Note that  the ratio $n_0/(\pi b^2)$ has the dimension of a volume density. 

In the short wavelength limit   $kb\gg1 $ these disk  waves become Langmuir waves with frequency $\omega_{Dpe}$,    but with a small group  velocity  ($ {\cal O} (1/(k b)^2)$ that does not depend on the disk temperature.  In the long wavelength limit we find  (logarithmically corrected) ``cold electron disk sound waves''   {with dispersion relation }of the form 
\begin{equation} \label{17}
\omega^2 =  k^2 v^2_D\,  \big[C_1 + C_2\ln{(kb)^2}\big] \, , 
\end{equation}
where the velocity $v_D $ is defined by 
\begin{equation} \label{17b}
 v^2_D  = b^2 \omega_{Dpe}^2= \frac {4 Q^2 n_o}{M_e}, 
\end{equation}
and $C_1$ and $C_2$ are  the  constants defined in  ( f{Vexp}). A plot  of  the normalized frequency $\omega/\omega_{Dpe}$,  and  of the group velocity is shown in Figure  f{cold}.   We recall that in a   Coulomb  plasma there is no  velocity directly corresponding to $v_D$ and thus no  cold electron-disk sound waves.

\begin{figure} [h!]\center
\includegraphics[width=.49\textwidth]{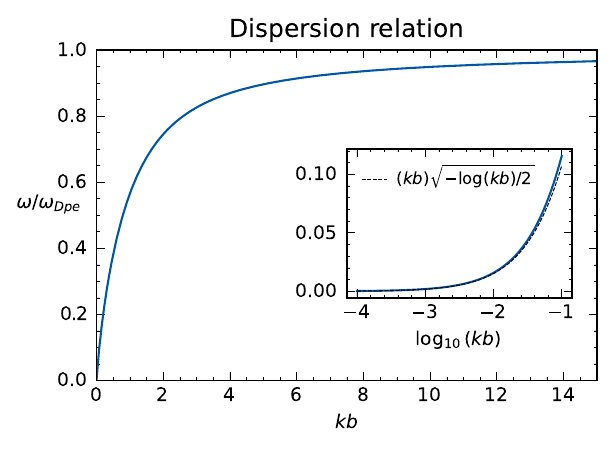}
\includegraphics[width=.49\textwidth]{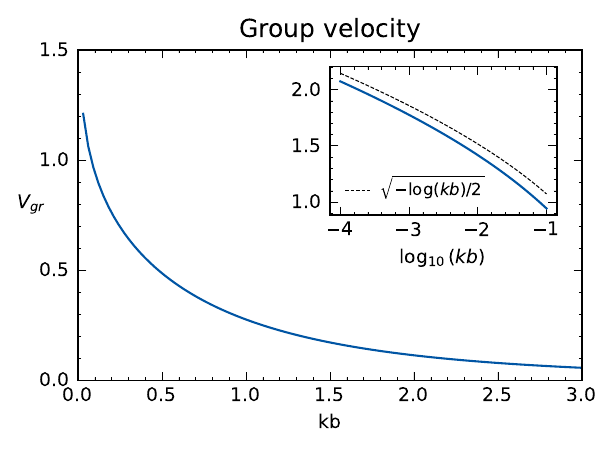}
\caption{Plots of  the normalized frequency $\omega/\omega_{Dpe}$ (left frame)   and of the group velocity (right frame)  versus $kb$. Note their   {square-root logarithmic behaviour  for $kb \to 0$.} }\label{cold}
\end{figure}

\subsection{Vlasov  equation and  the contribution of resonant disks} \label{dynamwarm}

The treatment of the Vlasov kinetic equation  for a disk plasma follows standard lines with two minor differences: the expression of the interaction potential $V(\xi)$ is different from the Coulomb potential and it may be  more convenient to use the disk charge  density  as the independent variable instead of the  electrostatic potential.

Thus we have the integro-differential equation 
\begin{equation} \label{18}
\frac{\partial { f}_j(x,v,t)}{\partial t} + v \frac{\partial {f}_j(x,v,t)}{\partial x} + \frac{{{\cal F}}_j(x,t)}{M_j} \frac{\partial { f}_{j}(x,v,t)}{\partial v} = 0, \end{equation}
where $ {f}_{j}(x,v,t)$ is the  phase space distribution function of the $j$ species ($j= e, i$), $v$ is the disk velocity  coordinate in phase space and   ${\cal F}_j(x,t) $ is the electric force, see  ( f{7}), due to the disk  charge density
\begin{equation} \label{19}
 {\rho}(x,t)  =  Q \left[\int_{-\infty}^{+\infty} \!\! dv\, { f}_{i}(x,v,t) \, - \, \int_{-\infty}^{+\infty}   \!  \! dv \,  { f}_{e}(x,v,t) \right] ,
\end{equation}
where  the electron and ion disk number densities are given by
\begin{equation} \label{19b}
 n_{i,e}(x,t)  =  \int_{-\infty}^{+\infty} \!  \! dv \,{ f}_{i,e}(x,v,t)\, .
\end{equation}
In a homogeneous disk plasma the linearized Vlasov  equation takes the form
\begin{equation} \label{18bis}
\frac{\partial {\tilde f}_j(x,v,t)}{\partial t} + v \frac{\partial {\tilde f}_j(x,v,t)}{\partial x} + \frac{{\tilde {\cal F}}_j(x,t)}{M_j} \frac{\partial { f}_{j0}(v)}{\partial v} = 0 , 
\end{equation}
where $ { f}_{j0}(v)$  and ${\tilde f}_j(x,v,t)$  are the equilibrium and  the perturbed distribution functions and  $\tilde {\cal F}_j(x,t) $ is  the perturbed force.

Assuming  immobile ion disks and   wave perturbations of the form
\begin{equation} \label{20}  
{\tilde f}_{e}(x,v,t) = {\tilde f}_{0}(v)  \exp {[-i (\omega t -  kx)]}, \quad 
{\tilde \rho}(x,t) =  {\tilde \rho}_0\, \exp {[-i (\omega t -  kx)]} , 
  \end{equation}
 with 
 \begin{equation} \label{20b}  
  {\tilde \rho}_0 = -Q   {\tilde n}_{e0} =  -Q \int_{-\infty}^{+\infty}   \!  \! dv \,   {\tilde f}_{e0}(v) \,, 
   \end{equation}
we  obtain the dispersion relation
\begin{align} \label{21}
 1 &= - 4  k Q^2   \frac{{\hat {V}}(kb)}{M_e} \int_{-\infty}^{+\infty}   \!  \! dv \,  \frac{ \partial {f}_{e0}(v)/\partial v}{\omega - k v} \,
 \nonumber\\
 &= -  \frac{\omega_{Dpe}^2}{k}  [(kb)^2{\hat {V}}(kb)] \int_{-\infty}^{+\infty}   \!  \! dv \,  \frac{ \partial [{f}_{e0}(v)/n_0]/\partial v}{\omega - k v}  \,.
 \end{align}
The velocity integral  in  ( f{21}) has to be taken according to the Landau prescription and, as for charged  particles in a Coulomb plasma, resonant disks satisfy the condition $\omega = k v$.

 In the cold limit  when $\omega/k \gg v_{ {\text{th}e}}$, with $v_{ {\text{th}e}}$ the  thermal velocity defined  as 
the halfwidth of the distribution function ${ f}_{e0}$,  ( f{21})  becomes
 \begin{equation} \label{22}
 1 =  4   Q^2  n_0  \frac{k^2}{\omega^2}  \frac{{\hat {V}}(kb)}{M_e} = \frac{\omega_{Dpe}^2}{\omega^2}  \big[(kb)^2{\hat {V}}(kb)\big] ,\end{equation}
 which coincides with  ( f{16}).  The dispersion relation ( f{21}) depends on  the normalized wave number $kb$ and in addition on   the equilibrium dimensionless parameter 
\begin{equation} \label{23-0}
 \Theta = (v_{ {\text{th}e}}/v_D)^2 \,  , 
 \end{equation}
that can be interpreted as the dimensionless electron disk  ``temperature''  $\Theta =   T_e/(4 Q^2 n_0)$ or, equivalently,  as  the square of the ratio between  the  disk  Debye length $ \lambda_{Dpe}$ 
defined as,
\begin{equation} \label{23-1}
 \lambda_{Dpe}=  v_{ {\text{th}e}} /\omega_{Dpe} \,, 
 \end{equation}
and the disk radius $b$,  (see  Section  f{screening}). For disk Langmuir  waves the cold limit $\omega/k\gg v_{ {\text{th}e}}$  implies $v_{ {\text{th}e}}/v_D \ll 1/(kb)\ll1$, while  for cold sound waves it implies the  condition $v_{ {\text{th}e}}/v_D \ll |\ln{(kb)^2}|$ where  $|\ln{(kb)^2}| \gg 1$.

Normalizing velocities over $v_D$  the dispersion relation can be rewritten  in terms of $ {\bar \omega} =\omega/\omega_{Dpe}$, \, ${\bar v} = v/v_d$ and $\kappa = kb$ as 
\begin{equation} \label{23}
 1 = - \kappa{\hat {V}}(\kappa) \int_{-\infty}^{+\infty}   \!  \! dv \,  \frac{ \partial {\bar f}_{e0}({\bar v})/\partial {\bar v}}{{\bar \omega} - \kappa {\bar v}} \,,
\end{equation}
where $
 {\bar f}_{e0}({\bar v}) = v_D \, f_{e0}(vt )/n_0$  is  the dimensionless disk  distribution function.

The linear dispersion relation of electron disk waves,   as obtained by integrating  ( f{23}) numerically   for a Maxwellian electron disk distribution and immobile ion disks,  is  shown in Figure  f{KINET} while 
the  phase space trapping     of electron disks  in a finite amplitude wave  at $t\,\omega_{Dpe} =  190$ is shown in  Figure  f{rNL}.

 \begin{figure} 
\includegraphics[width=0.51\textwidth]{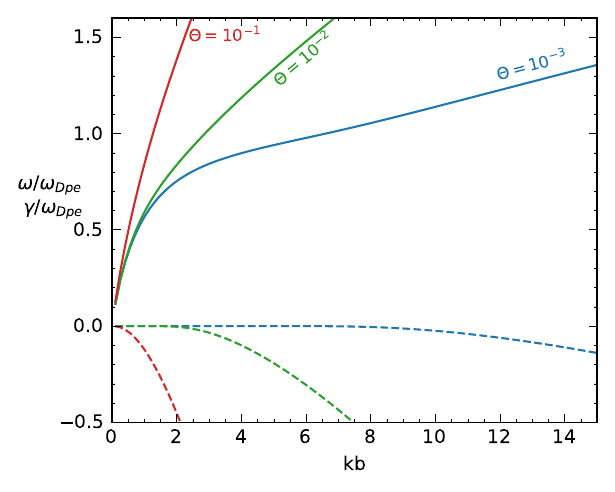}
\includegraphics[width=0.51\textwidth]{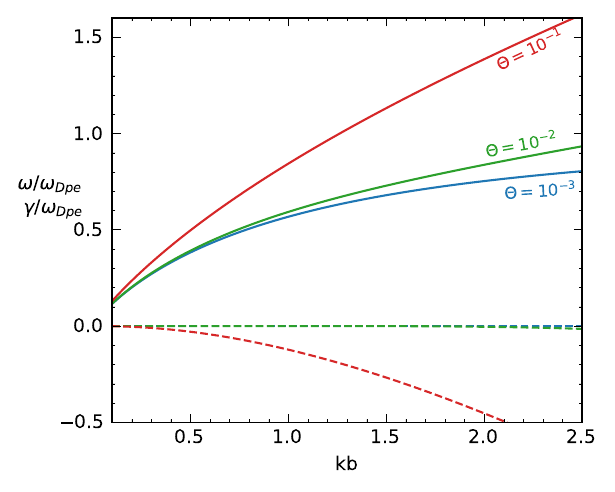}
\caption{Normalized frequency,   {solid lines},  and negative growth,   {dashed lines}  (Landau damping due to resonant disks) 
of electron disk waves versus $kb$ for different values of the  dimensionless temperature  $\Theta$. The right frame  presents   an enlarged plot that covers   small values  of $kb$ and  returns, for positive values approaching zero,  the   {square-root logarithmic} behaviour of cold waves. For  warm Langmuir waves the adiabatic index  $= 3$ as in a  1-D Coulomb plasma.  \label{KINET}}
\end{figure}

\begin{figure} [h!]\center
\includegraphics[width=0.85\textwidth]{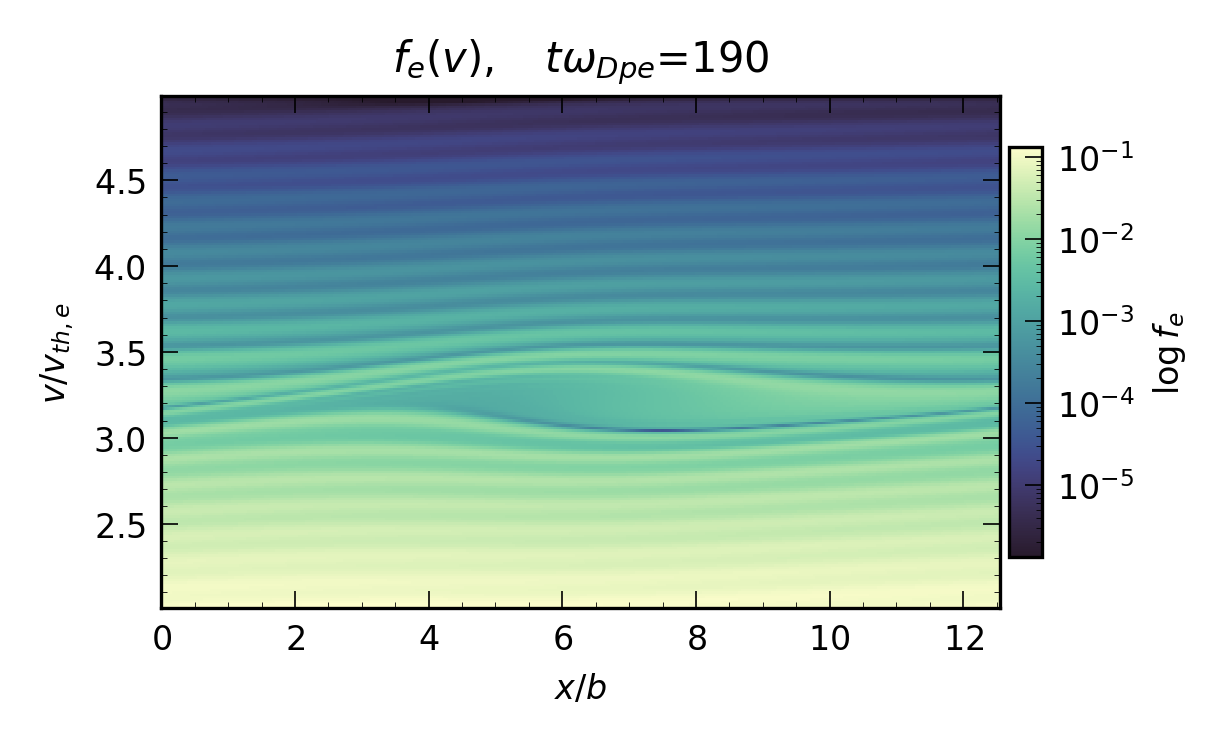}
\caption{Phase space of electron disks at the resonant velocity in a finite amplitude  electron disk plasma wave for:
$\Theta=0.2$, mobile ion disks  with  $M_i/M_e=1836$. The  amplitude of the initial  perturbation of the normalized density of the electron disks  is  ${{\tilde n}_e}/{n_0}  = 0.03 $ and $ kb=0.5$.
\label{rNL}}
\end{figure}

\subsection{Disk charge screening at thermodynamic equilibrium}\label{screening}

In order to  determine the screening of a stationary  test disk of charge $Q_{t}$  at $x=0$ produced  by the spatial rearranging of the other disks,   we write   the electron disk  density in terms of the screened potential $\varphi_{s}$ as 
 \begin{equation} \label{11}
n_e(x) =   n_e(0) \exp{ [(Q \varphi_{s} )/ T_e ] } \sim  n_e(0) [ 1+Q \varphi_{s} / T_e ] ,
\end{equation}
and neglect, for the sake of simplicity the perturbation of the density of the ion disks.
Note that in this disk model, contrary to a  3-D Coulomb plasma,  the approximation $Q \varphi_{s} / T_e \ll 1$ remains valid even at close distance.  Setting $n_i =   n_e(0) = n_0$, we can write the charge density of the  test disk,  plus the screening density  given by the second term in   ( f{11}),  as 
 \begin{equation} \label{12}
\rho (x) =   Q_t\delta(x) - Q^2   n_0 \varphi_{s} / T_e, 
\end{equation}
which when inserted into  ( f{6}) gives the integral equation
\begin{equation} \label{13}
\varphi_{s}(x) =   Q_t  (4/b) V(|x|/b) -  4(Q^2 n_0/T_e)  \int_{-\infty}^{+\infty}  \frac{ dx'}{b}\,  \varphi_{s}(x')V(|x-x'|/b)\,.
\end{equation}
Since the integral term  in  ( f{13}) is a convolution product,  ( f{13}) can be solved by performing a Fourier transform with respect to the $x$ variable (See   ( f{Vfur})  and Appendix  f{App1} for analytical details of the functions involved).  Note that the factor in front of the integral is dimensionless  and can be used to define  the disk Debye length   $\lambda_{Dpe}$ 
setting $4(Q^2 n_0/T_e) = 1/\Theta = b^2/\lambda_{Dpe}^2$ (see  ( f{23-1})).

The Fourier transform of  ( f{13}) reads
\begin{equation} \label{F13}
{\hat \varphi}_{s}(kb) =   Q_t  (4/b) {\hat V}(kb) -  4  (Q^2 n_0/T_e) {\hat \varphi}_{s}(kb)\,  {\hat V} (kb)\,,  
\end{equation}
i.e., 
\begin{equation} \label{F13b}
{\hat \varphi}_{s}(kb)   =   \frac{Q_t  (4/b) {\hat V}(kb)}{ 1 + 
4  (Q^2 n_0/T_e)   {\hat V} (kb)]}  \leq 
Q_t  (4/b) {\hat V}(kb)\,, 
\end{equation}
where we define ${\hat \varphi}_s$  as 
\begin{align} \label{FF13}
& {\hat \varphi}_{s}(kb) = \frac{1}{(2\pi)^{1/2}} \int_{-\infty}^{+\infty}  \frac{dx}{b} \, \varphi_s(x) \, \exp{ {\{-i[(kb)\, (x/b)]\}}}\, .
\end{align} 
Inverting the Fourier transform and defining  $\Phi_{s}(X)  = (b/4)\, \varphi_s(x) $,
from  ( f{F13b}) we obtain in dimensionless variables  with $X = x/b$ 
\begin{equation} \label{FF13b}
 \Phi_{s}(X)  = \frac{2 Q_t }{(2\pi)^{1/2}}\int _{0}^{+\infty}\!\! {d\kappa} \, \frac{{\hat V}(\kappa)}{1 +   {\hat V} (\kappa)/\Theta} \, \,  \cos{(\kappa X)}\, .
\end{equation} 
It is not direct  to perform  the integration in  ( f{FF13b})  analytically  because   of the logarithmic  dependence of  ${\hat V}(kb)$ for  $k b\ll 1$ (see  ( f{Vexp})), which leads  to a non-exponential  decay at large distances of the screened potential that  is not present in a 3-D Coulomb plasma.  This non-exponential  decay   is confirmed by  integrating  ( f{FF13b}) numerically for different values of 
the dimensionless  temperature $\Theta$. 

 {In Figure  f{HW} we  show   the screened electric  field  $E_{\rm scr}$,  as defined by  differentiating  ( f{FF13b}) with respect to $X$,
for $\Theta = 2$.  We see that 
 for large values of $X$,  the screened electric field   has a power  law  behaviour $ \propto X^{-2.2} $  and decays faster than the unscreened field. }

   \begin{figure}[h!]
\center
\includegraphics[width=0.61\textwidth]{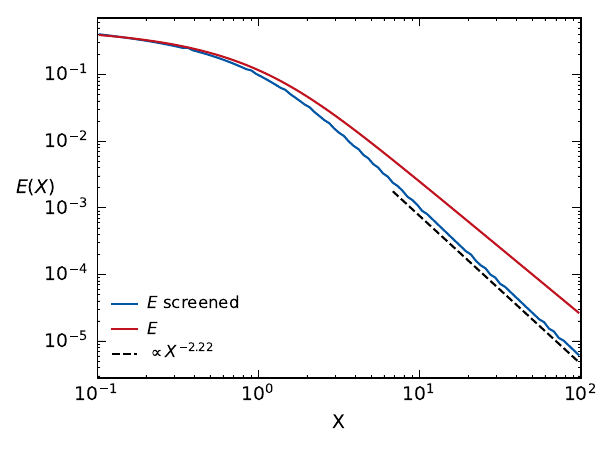}
\caption{ {Log-Log plot of  the  (normalized)  screened  electric  field  (blue) drawn against  the unscreened field   (red) for $\theta = 2$.  The power law behaviour of the   screened  electric  field is noted explicitly (black dashed line)}}
\label{HW}
\end{figure} 

\subsection{Plasma Parameter}\label{g-Parameter}

Disregarding  corrections due to   the presence of the  non-exponential decay of  the screened potential, 
we define the plasma parameter $g$ for a  disk plasma  by referring  to the length  $ \lambda_{Dpe}$ in  ( f{23-1})
as
\begin{equation}
g ^{-1} =  n  \lambda_{Dpe}= n \, \frac{v_{ {\text{th}e}}}{ \omega_{Dpe} }= (n b /2)^{1/2} \,
\left[\frac{ (M_e v_{ {\text{th}e}}^2/2)}
 {(Q^2/b)}\right] ^{1/2} , 
 \end{equation}
where it is assumed that $g \ll 1$.  Here $n ^{-1}$ plays the role of the  average distance between disks  while  $(M_e v_{ {\text{th}e}}^2/2 )/(Q^2/b)$ gives the ratio between the disk  kinetic energy  and a reference value of the disk-disk  interaction energy. 
The plasma parameter $g$ can be expressed in terms of  the disk radius $b$ and   the dimensionless temperature $\Theta = T_e /[4(Q^2 n_o)] $ as 
\begin{equation}
g ^{-1} =  n  \lambda_{Dpe} = n  b \,  \Theta^{1/2}\, .
\end{equation}

\section{BBGKY  hierarchy  for a disk plasma } \label{Numer}

In this section we apply the results obtained for the disk electrostatics to the  system of equations of the  BBGKY  hierarchy   in  Section  f{BBGKY} .

We insert  the disk  interaction potential  defined by  ( f{2}) into  the  time evolution equation (see   ( f{BB5})), for the one point distribution function  $ {F^{(1)}}(x_1,v_1,t) $ involving the two point distribution  function  $ {F^{(2)}} (x_1,v_1,x_2,v_2, t)$ 
and write for $N\gg 1$, where $N$ is the total number of disks,
\begin{align} \label{24} 
&\frac{\partial {F^{(1)} }(x_1,v_1,t)}{\partial t} + v_1\frac{\partial { {F^{(1)} }}(x_1,v_1,t)}{\partial x_1}
  \\ 
  & \hspace{.75cm}
-\frac{N}{M_1}\frac{4 Q_1 Q_s}{ b}\,  \frac {\partial}{\partial v_1} \int dx_s \frac{  \partial V(|x_1-x_s|/b)}{\partial x_1}\, \int dv_s \,{F^{(2)}} (x_1,v_1,x_s,v_s, t) =0\,,
 \nonumber
\end{align}
where 
\begin{equation}  \label{FF2} 
 F^{(2)} (x_1,v_1,x_s,v_s,t) =  F^{(1)} (x_1,v_1,t) F^{(1)} (x_s,v_s,t) +   \Delta F^{(2)} (x_1,v_1,x_s,v_s,s) \,. 
 \end{equation}
Here, the electron and  the ion disk  distribution functions  have not been separated for the sake of notational compactness, but indices have been added to the charge $Q$ and to the mass $M$ depending on the charge and on  the mass of the disks involved.  In addition $x$ - $v$ coordinates with disk indices are used in the 2-D disk phase space  of each disk.

Similarly, the evolution  equation for  ${F^{(2)} }(x_1,v_1,x_2,v_2, t)$ in terms of  the three point distribution function ${F^{(3)}} (x_1,v_1,x_2,v_2,x_3,v_3, t)$ (see   ( f{BB6})),  becomes 
\begin{align}  \label{long} 
& \frac{\partial  {F^{(2)}}(x_1,v_1,x_2,v_2,t)}{\partial t} + v_1\frac{\partial  {F^{(2)}}(x_1,v_1,x_2,v_2,t)}{\partial x_1} 
+ v_2\frac{\partial  {F^{(2)}}(x_1,v_1,x_2,v_2,t)}{\partial x_2}  \\&
 -\frac{1}{M_1}\frac{4 Q_1Q_2 }{b}\,   \left[ \frac{\partial V(|x_1-x_2|/b)}{\partial x_1}\, \frac{\partial  {F^{(2)}}(x_1,v_1,x_2,v_2,t)}{\partial v_1} 
\right]  \nonumber \\ &  -\frac{1}{M_2}\frac{4 Q_1Q_2 }{b}\,   \left[ \frac{  \partial V(|x_1-x_2|/b)}{\partial x_2}\, \frac{\partial  {F^{(2)}}(x_1,v_1,x_2,v_2,t)}{\partial v_2} \right]  \nonumber 
 \\&  { -\frac{N }{M_1}\frac{4 Q_1 Q_s }{b}\,\left[  \frac {\partial}{\partial v_1} \int  dx_s  \frac{\partial V(|x_1-x_s|/b)}{\partial x_1}\, \int  dv_s \, {F^{(3)}} (x_1,v_1,x_2,v_2,  x_s,v_s,t)  \right]  } \nonumber  \\& 
 { -\frac{N }{M_2}\frac{4 Q_2 Q_s }{b}}\, {\left[  \frac {\partial}{\partial v_2} 
\int  dx_s \frac{\partial V(|x_2-x_s|/b)} {\partial x_2}\, \int  dv_s \, {F^{(3)}} (x_1,v_1,x_2,v_2,  x_s,v_s,t) \right ]}=0\,. \nonumber \end{align}
Clearly  ( f{long}) implies  ( f{24}) as can be seen by setting  $N-2 \sim N-1 \sim N$ and integrating over  $x_2$  and $ v_2$.  
Recalling the expansion in   ( f{BB8}) we rewrite 
 \begin{align}\label{2b}  
 &{F^{(3)}} ( x_{1}, x_{2},  x_{s}, v_{1} , v_{2}, { v}_{s},t) =  {  {F^{(1)}} (x_{1}, v_{1},t)  {F^{(1)}}(  x_{2}, v_{2},t) {F^{(1)}}(x_{s}, v_{s},t)}
 \\
  & \quad +  {  {F^{(1)}} (x_{s}, v_{s},t) \,   {\Delta  F^{(2)} } (x_{1},v_{1}, x_{2}, v_{2},t)}   +  {F^{(1)}}(x_{1}, v_{1},t) \,   {\Delta  F^{(2)} } (x_{2},v_{2}, x_{s}, v_{s},t) \   
\nonumber \\
&\quad +   {F^{(1)}} (x_{2}, v_{2},t)  \,  {\Delta  F^{(2)} } (x_{1},v_{1}, x_{s}, v_{s},t)  +    {\cal O} (g^2)  \nonumber    
\end{align}  
 in the form 
  \begin{align}\label{2c}  
   {F^{(3)}} ( x_{1}, x_{2},  x_{s}, v_{1} , v_{2}, { v}_{s},t)  &=  {  {F^{(2)}}( x_{1}, x_{2}, v_{1} , v_{2} ,t)     {F^{(1)}}(x_{s}, v_{s},t) } 
   \\
   & + {F^{(1)}}(x_{1}, v_{1},t) \,   {\Delta  F^{(2)} } (x_{2},v_{2}, x_{s}, v_{s},t)  
   \nonumber\\
   &+
 {F^{(1)}} (x_{2}, v_{2},t)  \,  {\Delta  F^{(2)} } (x_{1},v_{1}, x_{s}, v_{s},t)  +    {\cal O} (g^2)\, ,\nonumber
    \end{align}
 where the first and the second terms on the r.h.s. of  ( f{2b}) have been recombined into the first term   on the r.h.s. of  ( f{2c}).  
This   reorganization  of the terms is convenient  since  the disk marked by $s$ 
plays  a different role  than the disks marked as $1$ and $2 $  in ( f{long}).

\subsection{Closed system of equations}\label{closed}

We refer now to  a disk plasma in a  finite  spatial  linear domain of length $L$,  corresponding to the finite volume in  Section  f{BBGKY}  for a 3-D configuration,  and, in order to connect to the notation used in  Section  f{dynamwarm} for the Vlasov equation, we introduce an average linear disk density
 \begin{equation}
{\bar n} = N/ L\, ,
\end{equation}
and define
\begin{align}
f^{(1)} (x_1,v_1,t) &=  {\bar n} \,  F^{(1)}(x_1,v_1,t) \,,\\
  f^{(2)} (x_1,v_1,x_2,v_2,t)  &= {\bar n} ^2 F^{(2)} (x_1,v_1,x_2,v_2,t) \,.
\end{align}
Then   ( f{FF2}), with $s \to 2$,  becomes 
\begin{equation}  \label{repeat}
 f^{(2)} (x_1,v_1,x_2,v_2,t) =  f^{(1)} (x_1,v_1,t) f^{(1)} (x_2,v_2,t) +  \Delta f^{(2)} (x_1,v_1,x_2,v_2,t) \, . 
 \end{equation}

In terms of  the functions $f^{(1)} (x_1,v_1,t) $  and  $  f^{(2)} (x_1,v_1,x_2,v_2,t) $, 
  ( f{24}) and  \eqref{long} can be rewritten as 
\begin{equation}  \label{f1}
 \left[\frac{\partial  }{\partial t} + v_1\frac{\partial }{\partial x_1}   \right]  f^{(1)}(x_1,v_1,t) =  {\cal S}^{(1)} ( f^{(2)} ) \,,
\end{equation}
and
\begin{equation}  \label{f1b} 
 \left[\frac{\partial  }{\partial t} + v_1\frac{\partial }{\partial x_1}  
+ v_2\frac{\partial  }{\partial x_2} \right]  f^{(2)}(x_1,v_1,x_2,v_2,t)=  {\cal S}^{(2)} ( f^{(1)}, f^{(2)}, \Delta f^{(2)})\, .
\end{equation}
Here
\begin{equation}
{\cal S}^{(1)} ( f^{(2)} )  =
\frac{L}{M_1}\frac{4 Q_1 Q_s}{ b}\,  \frac {\partial}{\partial v_1} \int_{-L/2}^{+\:L/2} \!\! dx_s \,\frac{  \partial V(|x_1-x_s|/b)}{\partial x_1}\, \int_{-\infty}^{+\infty}  \!\!dv_s \,{f^{(2)}} (x_1,v_1,x_s,v_s, t)\, ,
\end{equation}
and
 \begin{align} & \label{verylong} 
{\cal S}^{(2)} ( f^{(1)}, f^{(2)}, \Delta f^{(2)} )  \, =  \, \frac{1}{M_1}\frac{4 Q_1Q_2 }{b}\,   \left[ \frac{\partial V(|x_1-x_2|/b)}{\partial x_1}\, \frac{\partial  {f^{(2)}}(x_1,v_1,x_2,v_2,t)}{\partial v_1} 
\right]  \\ &  
+ \frac{1}{M_2}\frac{4 Q_1Q_2 }{b}\,   \left[ \frac{  \partial V(|x_1-x_2|/b)}{\partial x_2}\, \frac{\partial  {f^{(2)}}(x_1,v_1,x_2,v_2,t)}{\partial v_2} \right]  \nonumber 
 \\&   {+\frac{L }{M_1}\frac{4 Q_1 Q_s }{b}\,\left[  \frac {\partial}{\partial v_1}  f^{(2)}(x_1,v_1,x_2,v_2, t)\int_{-L/2}^{+L/2} dx_s  \frac{\partial V(|x_1-x_s|/b)}{\partial x_1}\,
  \int_{-\infty}^{+\infty}dv_s \, {f^{(1)}} ( x_s,v_s,t)  \right]  } \nonumber  \\& 
{+\frac{L }{M_2}\frac{4 Q_2 Q_s }{b}\,\left[  \frac {\partial}{\partial v_2}  f^{(2)}(x_1,v_1,x_2,v_2, t)\int_{-L/2}^{+L/2} dx_s  \frac{\partial V(|x_2-x_s|/b)}{\partial x_2}\,
  \int_{-\infty}^{+\infty} dv_s \, {f^{(1)}} ( x_s,v_s,t)  \right]  } \nonumber  \\&    { +\frac{L }{M_1}\frac{4 Q_1 Q_s }{b}\,\left[  \frac {\partial}{\partial v_1} \int_{-L/2}^{+L/2} dx_s  \frac{\partial V(|x_1-x_s|/b)}{\partial x_1}\, 
 f^{(1)}(x_1,v_1,t) \int_{-\infty}^{+\infty} dv_s \, {\Delta f^{(2)}} (x_2,v_2,  x_s,v_s,t)  \right]  } \nonumber  \\& 
 { +\frac{L }{M_2}\frac{4 Q_2 Q_s }{b}}\, f^{(1)}(x_1,v_1,t) {\left[  \frac {\partial}{\partial v_2} 
\int_{-L/2}^{+L/2} dx_s \frac{\partial V(|x_2-x_s|/b)} {\partial x_2}\, \int_{-\infty}^{+\infty} dv_s \, {\Delta f^{(2)}} (x_2,v_2,  x_s,v_s,t) \right ]} \nonumber \\&  { +\frac{L }{M_1}\frac{4 Q_1 Q_s }{b}\, f^{(1)} (x_2,v_2,t) \left[  \frac {\partial}{\partial v_1} \int_{-L/2}^{+L/2} dx_s  \frac{\partial V(|x_1-x_s|/b)}{\partial x_1}\, \int_{-\infty}^{+\infty} dv_s \, {\Delta f^{(2)}} (x_1,v_1, x_s,v_s,t)  \right]  } \nonumber  \\& 
 { +\frac{L }{M_2}\frac{4 Q_2 Q_s }{b}}\, {\left[  \frac {\partial}{\partial v_2} 
\int_{-L/2}^{+L/2} dx_s \frac{\partial V(|x_2-x_s|/b)} {\partial x_2}\, f^{(1)} (x_2,v_2,t) \int_{-\infty}^{+\infty} dv_s \, {\Delta f^{(2)}} (x_1,v_1,  x_s, v_s,t) \right ]} 
. \nonumber \end{align}
Equations ( f{repeat}) and  \eqref{verylong} are a closed system of equations for the functions $f^{(1)}$, $f^{(2)} $ and $\Delta f^{(2)} $.

This system  can be solved by assigning at $t =0$ the initial values of  $f^{(1)}$ and   $\Delta f^{(2)} $.  \\ This   makes it possible to compute \begin{equation} \label{t0}  f ^{(2)} (x_1,v_1,  x_2, v_2, \, 0)  =  f^{(1)} ( x_{1},  v_{1}, \, 0) \,  f^{(1)} ( x_{2},  v_{2}, t=0)  + \Delta f^{(2) }(x_1,v_1,  x_2, v_2, \, 0)\end{equation} 
algebraically.\, \,    {The condition}
 \begin{equation} \label{consist} \int_{L/2} ^{L/2}  d x_2   \int_{-\infty} ^\infty d v_2 
 \, f^{(2)} ( x_{1}, x_{2},   v_{1} , v_{2},t)   =  {\bar n} \,  f^{(1)} ( x_{1},  v_{1}, t)\,, 
 \end{equation}
 i.e.,
  \begin{equation} \label{consist2}
   \int_{L/2} ^{L/2}  d x_2   \int_{-\infty} ^\infty \!\!d v_2 \,
 \,  \Delta f^{(2)} ( x_{1}, x_{2},   v_{1} , v_{2},t)   = 0
 \end{equation}
  { is valid by definition at  all times and thus the initial choice of $\Delta f^{(2)} (x_1,v_1,  x_2, v_2, \, 0 )$  must satisfy it.}
 
 {For $t>0$,  the functions   $f^{(1)} ( x_{1}, v_{2},t)$  {and} $f^{(2)} ( x_{1}, x_{2},   v_{1} , v_{2},t)$  can be advanced in time  using   \eqref{f1} and    \eqref{f1b} respectively and calculating  at every step  $ \Delta f^{(2)} ( x_{1}, x_{2},   v_{1} , v_{2},t)$  algebraically. Since \eqref{f1} and    \eqref{f1b} are integrated independently, one must  verify  that they  satisfy  the   consistency condition ( f{consist})  for all times.  This check provides  a tool  that will be used in order to test the accuracy of their  numerical integration.\\
 Conversely,  one can avoid to advance in time both \eqref{f1} and   \eqref{f1b}  and   use  ( f{consist})  to transform equation  \eqref{f1b}  for $ f^{(2)} ( x_{1}, x_{2},   v_{1} , v_{2},t)$ into  an integro-differential equation by expressing  $f^{(1)}$ in terms of $f^{(2)}$ by direct  space and velocity integration. In this approach a check of the accuracy of the numerical integration will be the conservation of the total number of particles}.

 {By integrating \eqref{f1} and    \eqref{f1b}  it will be possible to ascertain  directly whether, and under which conditions, 
  $ \Delta f^{(2)}
({x}_{1}, {v}_{1},   {x}_{2}, {v}_{2},t)   $
does reach an asymptotic form that   is uniquely determined by the instantaneous form  of  $f^{(1)}( {x}, { v}, t)$    i.e.  whether  we can write (see \eqref{BB9})
   \begin{equation}\label{BB91}   
   {  \Delta f^{(2)}}({x}_{1}, {v}_{1},   {x}_{2}, {v}_{2},t) = { \Delta f^{(2)}}[f^{(1)}( {x_1}, {v_1}, t), f^{(1)}( {x_2}, {v_2}, t)]\, ,
   \end{equation}
and determine how the characteristic time that may be  needed to reach this asymptotic form compares with the dynamical time scale of the system.}

\section{Generalizations and future developments}\label{fut}
  
The model  plasma  introduced in the present article can  also provide interesting  information when applied  to problems different from the closure of the BBGKY hierarchy for  an  electrostatic system.

In a direct generalization   one can  conceive  of a  plasma  of gravitationally interacting disks with  different masses  (light and heavy disks) by suitably modifying the  coefficients and the sign  in the  expression of the interaction energy  in  ( f{1}).
 Such a modification would be of interest for dealing with the problem of the mass segregation in 1-D stellar system \cite{bertin}  representing, e.g.,  a   spatially flat  and uniform  distribution  of stars. Mass segregation in  star systems   cannot be described within the  (mean field) Vlasov  framework, where the star trajectories do not depend on their mass, but  are  allowed if a collision operator, arising from star correlations, is included. For such a system the model presented in this article would  make it possible to address the problem of the mass segregation  by deriving a collision operator in a 1-D configuration while still maintaining that the potential vanishes  at infinity.   Simulations with a large but finite number of gravitationally interacting disks  can be easily performed  and show disk evaporation (i.e., disks becoming unbound), mass segregation, and  the  disk phase space behavior for  the  different disk masses \cite{moruzz}.

Returning to the  problem of the validation of the Bogoliubov assumption,
the system of  ( f{f1}) and  \eqref{f1b}  will be integrated using an Eulerian Vlasov code \cite{Mang} that solves the Vlasov equation in a  multidimensional phase space. 
The advantage of the Eulerian approach \cite{Calif}  is that it provides an almost zero-noise  integration  procedure,   even at  the smaller scales in a fully non-linear, turbulent regime; however, at the expense of a huge computational cost.  The original version  of the code must be adapted to a larger phase space but the splitting approach 
of a multi-advection equation will be preserved. Periodic space boundary conditions will be imposed.
 With respect to the usual electrostatic  approach in plasma physics where the Vlasov equation  is 
coupled to the Poisson equation, here  the  electric potential  is obtained from the disk charge density by means of  a Green's function where the kernel is the  interaction potential between  individual disks. This procedure  has been already implemented  for the  numerical integration of the disk Vlasov equation that  led to the plots of the warm plasma dispersion relation in  Section  f{dynamwarm}.

The numerical solution of the system of  ( f{f1}) remains  a formidable challenge even on the  latest generation of GPU supercomputers.  We plan to  investigate  a   selected set  of physically relevant conditions   {with the aim of verifying  within the model introduced in this article   under which conditions   the Bogoliubov approach holds  and whether  a ``collision'' operator can be defined  under general plasma  conditions.}
\\
 { However dimensionality plays a major role in plasma dynamics, as is well known for example in the case of plasma turbulence. 
In a one-dimensional disk plasma  the   charged disk dynamics is more constrained  than that of charged particles in a real three-dimensional  plasma and  it is not possible to define and impact parameter.  In this sense one cannot  distinguish for example   between soft and hard ``collisions''.  
Thus the model introduced in this article cannot be expected to provide a conclusive   validation or  disproval of the  Bogoliubov assumption  in a three-dimensional plasma but can provide information on how the time separation that is at the basis of this assumption depends on the initial particle correlation, on the presence of distribution functions with highly energetic particle tails, of large electric field fluctuations or strong plasma spatial inhomogeneities.
These features can also occur in a three-dimensional plasma so that 
  information  that can be gained from  the model described in the present manuscript can  provide  useful indications  in order to identify  relevant processes that  can be at work in a  three dimensional plasma and possibly not be covered by the Bogoliubov assumption.}
\bmhead{Acknowledgements}

It is a pleasure to thank Prof.\ G.\  Moruzzi for  his help in the study  of the  discrete  multi-disk dynamics   for both   the  cases of  electrostatic and of gravitational interaction.  PJM acknowledges support from the US Department of Energy DE-FG02-04ER54742.


\begin{appendices}

\section{Disks electrostatics} \label{App1}

A set of relevant formulae  derived in   \cite{Disk} are listed  below with   some notational adaptations together with the derivation of  ( f{Vfur}).

The axially symmetric electrostatic potential $\varphi(r,x)$ created by a uniformly charged disk  with its center at the coordinate origin, with  total charge $Q$ and radius  $b$ can be written as 
\begin{equation} \label{A1}
\varphi(r,x) =    2 Q    \int_{0}^{+\infty} \!\!ds \, J_0(sr)\frac{J_1(sb)}{sb}\exp{(-s|x|)}\,,
\end{equation}
where $J_0$ and $J_1$  are  Bessel functions, while the $x$-component of the electric field can be written as
\begin{equation} \label{A2}
{\epsilon}_x (r,x) =    {{\rm sign ({\it x})}} \,  \frac{2 Q}{b}    \int_{0}^{+\infty} \!\!ds \, J_0(sr) J_1(sb) \, \exp{(-s|x|)}\,.
\end{equation}
At $r=0$, it reduces to 
\begin{align} \label{A3}
{\epsilon}_x (0,x) &=     {{\rm sign ({\it x})}} \,   \frac{2 Q}{b}    \int_{0}^{+\infty}\!\! ds \, J_1(sb) \, \exp{(-s|x|)}
\nonumber\\ 
&=   {{\rm sign ({\it x})}} \, \frac{2 Q}{b^2}  \left[ 1 - \frac{|x|/b}{(1 + x^2/b^2)^{1/2}}\right]\,.
\end{align}
The expression for the interaction energy in  ( f{1}) is given by 
\begin{equation} \label{A4}
{\cal W} (|x|) =    4  \frac{Q_1\,  Q_2}{b} V(|x|/b) =  \frac{4 Q_1Q_2}{b}    \int_{0}^{+\infty} \!\!ds\,  [J_1(s)/s]^2 \, \exp{(-s|x|/b)}\, ,
\end{equation}
which follows  by integrating  ( f{A1}) over the disk surface (see  (15) of   \cite{Disk}).

\bigskip 
\section{Fourier transform  of $V(\xi)$} \label{App2}

Using   ( f{A4}), the Fourier transform of $V(\xi)$  in  ( f{2}) can be  obtained as follows 
\begin{align} \label{A5} 
{\hat V} (kb) &=   \int _{-\infty}^{+\infty} \frac{dx}{b} \,V(|x|/b) \exp{ {\{-[i (kb) \, (x/b)]\}}}   
\nonumber\\ &  
=      \int_{0}^{+\infty} \!\! ds\,   [J_1(s)/s]^2  \, \int _{-\infty}^{+\infty} \frac{dx}{b}  \, \exp{(-s|x|/b)} \exp{ {\{-[i (kb) \, (x/b)]\}}} \nonumber \\
&=
2  \int_{0}^{+\infty} \!\!ds \, [J_1(s)/s]^2  \,   \int _0^{+\infty} \frac{dx}{b}  \, \exp{(-s x /b)} \cos{[(kb) \, (x/b)]}  \nonumber\\ 
&=
2  \int_{0}^{+\infty}\!\! ds\,  [J_1(s)/s]^2  \,  \frac{s}{(kb)^2 + s^2} = \frac{1}{  \pi^{1/2} (kb)^2}\, G^{2,2}_{2,4}\left((kb)^2\left| ^{1/2,1}_{1,1,-1,0} \right.\right)
\,,
\end{align} 
 where Meijer $G$ is the Meijer G function   \cite{Nist}.
 In the  notation  used by ``Mathematica''  as
  (see  \url{https://reference.wolfram.com/language/ref/MeijerG.html}) the Meijer G function  is denoted as
\begin{equation}
 G^{2,2}_{2,4}\left(z\left| ^{a_1,a_2}_{b_1,b_2,b_3 , b_4} \right) \right. = {\rm Meijer} G[
  \{\{a_1, a_2\},\{...\} \}, 
  \{\{b_1,b_2\},\{b_{3},b_{4}\}\},z]  \,.
  \end{equation}.

\end{appendices}

\section*{Declaration}

The authors declare that there are no conflicts of interest.

\end{document}